\documentstyle{elsart}
\makeatletter
\def\fps@figure{p}              
\makeatother
\begin{document}
\begin{frontmatter}


\title{A liquid-helium cooled large-area
  silicon PIN photodiode x-ray detector}


\author[ICEPP]{Yoshizumi Inoue},
\author[PHYS]{Shigetaka Moriyama},
\author[PHYS]{Hideyuki Hara},
\author[PHYS]{Makoto Minowa}, \and
\author[PHYS]{Fumio Shimokoshi}
\address[ICEPP]{International Center for Elementary Particle Physics,
  University of Tokyo, 7-3-1 Hongo, Bunkyo-ku, Tokyo 113, Japan}
\address[PHYS]{Department of Physics, Graduate School of Science,
  University of Tokyo, 7-3-1 Hongo, Bunkyo-ku, Tokyo 113, Japan}


\begin{abstract}
An x-ray detector using a liquid-helium cooled
large-area silicon PIN photodiode
has been developed
along with a tailor-made charge sensitive preamplifier whose
first-stage JFET has been cooled.
The operating temperature of the JFET
has been varied separately and optimized.
The x- and $\gamma$-ray energy spectra for an \nuc{241}{Am} source
have been measured with the photodiode operated at 13 K.
An energy resolution of 1.60 keV (FWHM) has been obtained
for 60-keV $\gamma$ rays
and 1.30 keV (FWHM) for the pulser.
The energy threshold could be set as low as 3 keV.
It has been shown that a silicon PIN photodiode serves as a
low-cost excellent x-ray detector which covers large area
at 13 K.
\end{abstract}
\end{frontmatter}


We have developed an x-ray detector to be employed
in our solar axion experiment.
The axion is a light pseudoscalar particle
introduced to solve the strong $CP$ problem \cite{axion}.
Sikivie \cite{Sikivie}
proposed an experiment to detect the axions
emitted by the sun using a system of
a strong magnetic field
and an x-ray detector, called the axion helioscope.
In the magnetic field, solar axions convert to x rays
of black body radiation spectrum
with an average energy of 4.2 keV.
The conversion can be enhanced by filling the conversion
region with dense gas \cite{Bibber}.
We are going to adopt cold helium gas
as conversion medium
with a temperature just above the boiling point
at one atmosphere.
In the experiment, an x-ray detector with the
following specification is needed:
\begin{itemize}
\item sensitive to x rays of 3--10 keV
\item energy resolution better than a few keV
\item operable at low temperature
\item sensitive area larger than 10 cm$^2$
\item low cost
\item extremely low background
\end{itemize}

Although gaseous detectors are often used for the soft x-ray detection,
they are unusable at low temperature.
By contrast, low temperature is a favorable environment for
semiconductor detectors since their leakage currents decrease
with temperature,
hence high energy resolution is achieved.
Among various semiconductor detectors,
Si (Li) detectors are commonly used for low energy photon detection,
but they are
very expensive so that they are not appropriate for a large-area
x-ray detector.

Silicon PIN photodiodes usually used as light detectors
are known to work also as good radiation detectors at liquid
nitrogen temperature \cite{PIN:LN2}.
They are commercially available at relatively low price in
various size and shapes.
The thickness of their active layers is
typically 200--500 $\mu$m, which is sufficient for the
detection of soft x ray,
but is so thin that silicon PIN photodiodes are almost
insensitive to $\gamma$ rays of higher energy.
The latter feature, the low sensitivity to the background
$\gamma$ rays,
makes silicon PIN photodiodes even advantageous over the expensive
Si (Li) detectors because lower background is expected for them.

In this letter, we present the result of a measurement
we have performed to see whether a silicon PIN photodiode
possesses the required performance,
namely,
sensitivity to the x rays
and enough energy resolution,
even at the temperature far below liquid-nitrogen temperature.


\begin{figure}
  \caption{The schematic view of the experimental setup}
  \label{fig:setup}
\end{figure}

The experimental setup is illustrated in Fig. \ref{fig:setup}.
Since electronic noise dominates
the energy resolution of the detector system,
the first-stage junction field-effect transistor (JFET) and the
feedback resistor of the charge sensitive preamplifier were cooled
together with the PIN photodiode
in order to reduce the thermal noise.
They were put into a vacuum vessel
and the vessel was immersed in a liquid-helium dewar.
An \nuc{241}{Am} source was also put inside the vessel.

A silicon PIN photodiode with a thickness of 500 $\mu$m,
Hamamatsu S3204-06, was used.
It is a windowless photodiode with an active
area of $\rm18\,mm\times18\,mm$
and is supplied on a ceramic base.
The typical capacitance value is specified
to be 80 pF at a reverse bias voltage of 100 V
by the manufacturer.
Between the source and the PIN photodiode is
a copper collimator of 1-mm thick and with
a hole of $^\phi$1.5 mm in the center.
In order to reduce the microphonic noise
the PIN photodiode was softly
supported with a phosphor bronze ribbon
to a liquid-helium cooled copper plate
soldered on the end cap.
Its temperature was monitored with a carbon resistance thermometer
(Allen-Bradley 56-$\Omega$ solid carbon resistor) stuck on its back.

Other electronic components,
such as the JFET and the feedback resistor,
were mounted on a glass-epoxy printed circuit board
and cooled with a 1-mm-thick copper-plate cold finger
fixed on the end flange at its base.
A low-noise JFET, Hitachi 2SK291, was used at the first stage
of the charge sensitive preamplifier.
Its transconductance $g_m$ has been measured to be
50 mS at room temperature
and 30 mS at liquid nitrogen temperature.
In order to obtain the optimum noise feature,
the operating temperature of silicon JFETs should be kept
at around 120--170 K \cite{Radeka}.
Therefore, the JFET was mounted on a copper chip
and a 6.5-mm-long stainless steel pipe
with outer diameter of 4 mm and thickness of 0.2 mm
was inserted between the copper chip
and the cold finger to keep the JFET at relatively high temperature.
In addition, a manganin wire was put around the JFET as a heater
and the JFET temperature was separately controlled
by varying the power fed to the heater.
The temperature of the copper chip was monitored with a
platinum resistor (Tama Electric SDT101A).

\begin{figure}
  \caption{The schematic description of the charge sensitive
    preamplifier. The PIN photodiode is directly coupled to
    the gate of the JFET.
    A serial 120-$\Omega$ resistor was inserted for the sake of phase
    correction.
    The feedback capacitance, shown in dotted lines, is the stray
    capacitance of the 5-G$\Omega$ resistor.}
  \label{fig:preamp}
\end{figure}

The schematic description of the charge sensitive preamplifier
is shown in Fig \ref{fig:preamp}.
The bias voltage fed to the PIN photodiode was set at 100 V, the value
which was recommended to produce full growth of the depletion layer.
The test input of the charge sensitive preamplifier was connected
to a pulser,
and the output signal of the preamplifier was fed
to a shaping amplifier (Camberra 2026),
then digitized with a peak-sensing ADC (Laboratory Equipment 2201A).


Before the vessel was cooled to liquid-helium temperature,
pulse shaping and JFET operating temperature were optimized
at liquid-nitrogen temperature.
To obtain the highest resolution for the pulser inputs,
the triangular pulse shaping was chosen
and the shaping time constant was fixed at 12 $\mu$s,
the longest time constant available with the shaping amplifier.
As the leakage current at this low temperature is negligible,
the serial noise is dominant. Thus it is quite natural that
the longest time constant yielded the best result \cite{Radeka}.
Then, to optimize the operating temperature of the JFET,
we measured the pulser resolutions
as a function of the measured JFET operating temperature
and searched for the minimum.
The optimum operating temperature of the JFET was
around 110--140 K,
which was consistent with the value shown in the ref. \cite{Radeka}.

\begin{figure}
  \caption{The typical x- and $\gamma$-ray energy spectrum measured
    for the \nuc{241}{Am} source at 13 K is shown.
    The operating temperature of the JFET was around 110--140 K.
    The pulser line is seen at 122 keV.
    Two $\gamma$-ray lines are distinguished at 26 keV and 60 keV,
    and three Np L x-ray lines at 14 keV, 18 keV, and 21 keV.
    The line at 8 keV is not originated from \nuc{241}{Am}, but is
    the fluorescent Cu K x-ray line due to the irradiation of
    \nuc{241}{Am} $\gamma$-rays on the copper collimator.}
  \label{fig:spectrum}
\end{figure}

Then the vessel was cooled down to liquid helium temperature.
The typical energy spectrum
for the \nuc{241}{Am} source measured with the
JFET at the optimum temperature is shown in Fig. \ref{fig:spectrum}.
The temperature of the PIN photodiode was $13\pm3$ K.
The 60-keV $\gamma$-ray line is clear, and
the 26-keV $\gamma$-ray line, the three Np L x-ray lines,
and the 8-keV Cu K x-ray line,
due to the fluorescence on the copper collimator,
are also distinguished in the spectrum.
The pulser resolution was 1.30 keV (FWHM),
and the energy resolution for the 60-keV $\gamma$-ray line was
obtained to be 1.60 keV (FWHM).
The latter was calculated only with the data of the higher
energy side of the peak as the 60-keV line shape is slightly
asymmetric around the peak due to scattered $\gamma$ rays.
The energy threshold could be set as low as 3 keV.
This is already a promising value for the detection of
axion-converted x rays with an average energy of 4.2 keV.

We also tried to lower the temperature of the PIN photodiode
by sticking it directly on the cooled copper plate,
and observed that the silicon PIN photodiode functioned
as an x-ray detector at 6 K,
but the microphonic noise was so severe that the measurement
of spectra was almost impossible with the simple method,
namely the combination of pulse shaping and pulse-height measuring.
Either isolation of the vibration itself or
a special discrimination of the microphonic noise
from the actual signal
is necessary to get rid of this problem.


In summary,
it has been demonstrated that a silicon PIN photodiode
functions as a good x- and $\gamma$-ray detector at 13 K.
With the charge sensitive preamplifier in which
the operating temperature of the JFET has been optimized,
the energy resolution for 60-keV $\gamma$ rays
has been obtained to be 1.60 keV (FWHM).
The energy threshold could be set as low as 3 keV.
The resolution is sufficient and the threshold is a promising
value for our requirements.

These values would be improved further if the large capacitance
of the PIN photodiode could be reduced,
for example, by using two half-sized
photodiodes and duplicating the electronics.
Such a setup will still be less expensive than to use Si (Li)
detector system.

Our next step is to overcome the microphonic problem.
The microphonic noise damages the energy spectra both
by broadening the peaks
and by raising the energy threshold
in an unreproducible manner.
If the microphonic noise could be eliminated,
the temperature frontier would be lowered further.


\begin{ack}
The authors thank Messrs. Takashi Uchihashi and Akinobu Kanda
for useful discussions on the cryogenic instruments.
\end{ack}



%
%
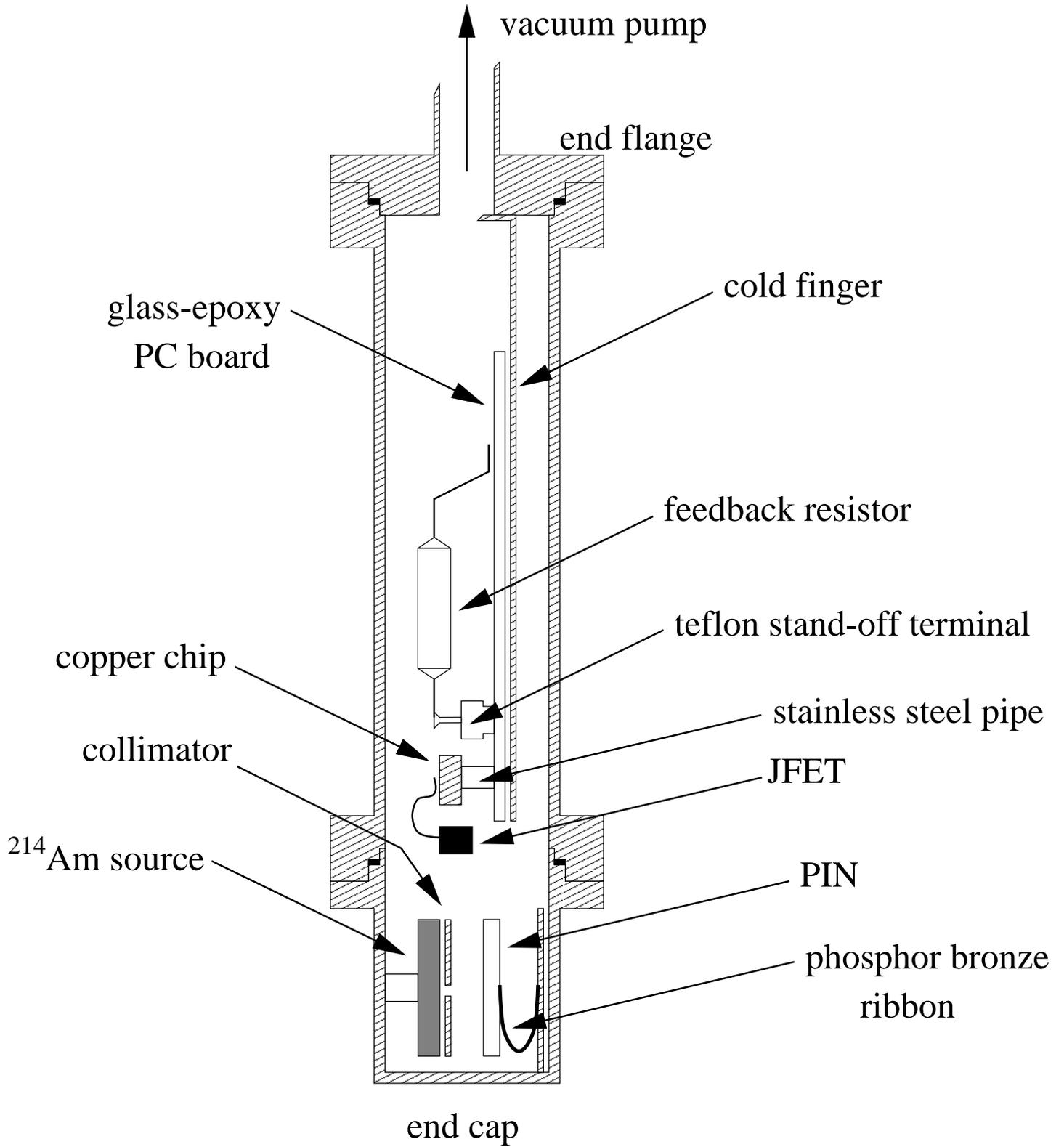
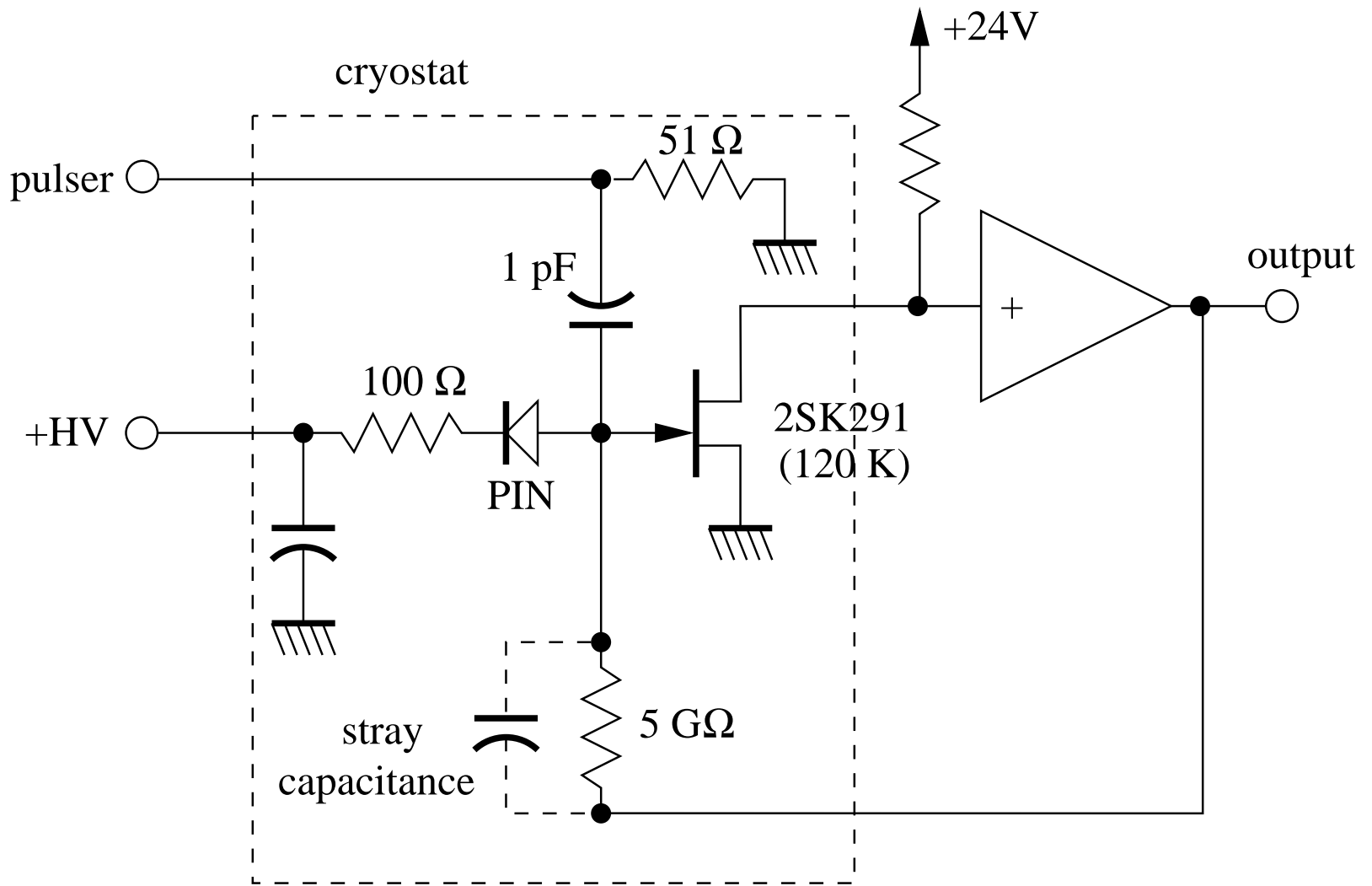
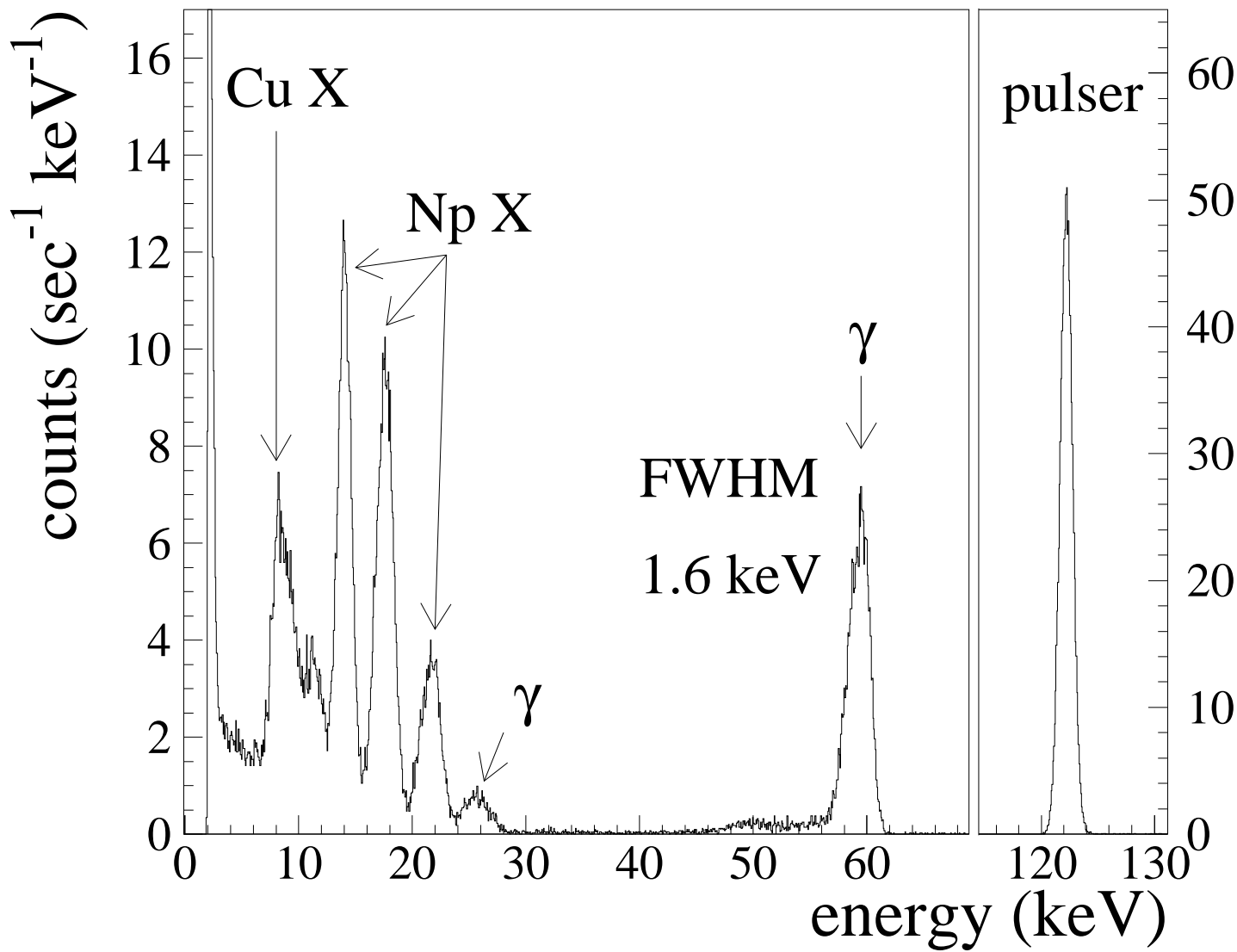
\end{document}